\documentclass[12pt]{iopart}
\begin{document}

\title[A comparative analysis of the electron energy distribution
function]{A comparative analysis of the electron energy
distribution function obtained by regularization methods and by a
least-squares fitting.}
\author{C. Guti\'{e}rrez-Tapia and H. Flores-LLamas}
\address{Departamento de F\'{\i}sica, Instituto Nacional de
Investigaciones Nucleares \\
A. Postal 18-1027, 11801 M\'{e}xico D. F., MEXICO}
\ead{cgt@nuclear.inin.mx}

\begin{abstract}
To establish the electron energy distribution function (EEDF), the
second derivative of a Langmuir probe current-voltage (I-V)
characteristic is numerically integrated using the Tikhonov
singular value decomposition regularized method. A comparison of
the numerically intagrated EDDF and by a least-squares fitting is
discussed. The used I-V characteristic is measured in an ECR
plasma source using a cylindrical probe and the plasma parameters
are determined by the Laframboise theory. This technique allows a
rapid analysis of plasma parameters at any gas pressure. The
obtained EEDF, for the case of the ECR plasma source, shows the
existence of two groups of electrons with different temperatures.
This result is associated with the collisional mechanism heating
taking place in ECR plasma sources, where low pressure plasma is
sustained by electron impact ionization of the ground state
molecules or atoms by energetic electrons arising in the resonance
zone.
\end{abstract}

\maketitle

\section{Introduction.}

The Langmuir probe is one of the simplest diagnostics tools of the
study of ionized gases. The probe usually consists of a small
sphere or a circular cylinder which is placed in the plasma at the
point of interest. Some external circuitry is provided so that the
electric potential of the probe can be varied. A plot of the total
current flowing between the probe and the plasma versus the probe
potential is called the current-voltage (I-V) characteristic.
Langmuir probes are highly effective in determining the electron
energy distribution function (EEDF), allowing both spatial and
temporal resolution. The EEDF is given by the second derivative of
the probe I-V characteristic.

In recent years, several groups have successfully used the
numerical differentiation of digital probe trace data. \cite{1, 2,
3}. Yet, numerically, the second derivative is a hard task and is
classified as an ill-posed problem. Ill-posed problems do not have
the properties of existence, uniqueness and stability \cite{4}.
One of the alternatives to solve this problem is the introduction
of the concept of conditional well-posed problems \cite{4}. On the
other hand, the problem of integration of the I-V characteristic
has evolved along with the solution of the Fredholm integral
equation of first class, which is obtained from the integration of
a second order differential equation. One of the most stable
methods to solve the Fredholm integral equation are the
regularization methods.

In order to characterize the processes occurring in plasmas it is
advisable to obtain the electron energy distribution function with
a high degree of accuracy. In particular, there has been much
interest in the electron cyclotron resonance (ECR) plasma sources
motivated by their applications in ionized physical vapor
deposition techniques (I-PVD) \cite{5}.

In this paper, the second derivative of a Langmuir probe
current-voltage (I-V) characteristic is numerically integrated
using the Tikhonov singular value decomposition (SVD) regularized
method in order to establish the EEDF as the most stable method.
The existence of two groups of electrons with different
temperatures, where low pressure plasma is sustained by electron
impact ionization of the ground state molecules or atoms by
energetic electrons arising in the resonance zone is shown. Also,
it is important to notice that the rate of convergence of the
Tikhonov method is much faster than that achieved with the
least-squares with an orthogonal decomposition (LSQR) and
truncated singular value decomposition (TSDV) methods. A
comparison between the EEDF obtained by the Tikhonov
regularization method and the one resulting from a least-squares
fitting is discussed.

\section{Basic equations.}

The electric potential profile near the probe can influence its
current collection by setting up ``barriers of effective
potential'' around the probe, thereby preventing some particles
from reaching it. This profile can be strongly influenced by the
space charge of the particles themselves. However, when
$r_p/\lambda_D$ is small enough (the sheath around the probe is
comparatively thick), this barriers disappears, and the current
collection is maximized. The currents collected under these
conditions then become the orbit-limited-currents. This assumption
is backed by the orbital motion limit (OML) theory \cite{6} which
implies a thick, collisionless sheath ($\lambda_e \gg r_p$). Thus,
the calculation of Laframboise \cite{6}, based on the more
complete theory of Berstein and Rabinowitz \cite{7}, showed that
the OML limit is achived for $\lambda_D \approx r_p$. Here $r_p$
is the probe radius (spherical or cylindrical), $\lambda_D$ is the
Debye radius, and $\lambda_e$ is the electron mean free path,
respectively.

When we assume the presence of a stopping field ($\lambda_e > r_p
> \lambda_D$), as demonstrated by Druyvesteyn in \cite{8, 9}, the second
derivative of the current $I$ with respect to the voltage $V$ of
the probe is proportional to the electron energy distribution
function as follows
\begin{equation}
F(\varepsilon) = n_e f(\varepsilon) = \frac{4}{eS}
\sqrt{\frac{mV_p}{2e}} \frac{d^2 I}{d V_p^2}, \label{1}
\end{equation}
where the energy is $\varepsilon=e(V_p-V_s)$, $V_p$ is the probe
voltage, $V_s$ is the plasma potential, $I$  is the electron
current to the probe, $S$ is the probe surface area, and $m,$ $e$
are the electron mass and charge, respectively. The contribution
of the ion current to the second derivative of the probe current
is normally insignificant, and it is therefore not subtracted in
these EEDF calculations.

The equation (\ref{1}) is in the form
\begin{equation}
\frac{d^2 I}{dV_p^2} = F(V_p), \label{2}
\end{equation}
where $F(V_p)$ is a function of the potential on the probe $V_p$
with the boundary conditions for the current given by
\begin{eqnarray}
I|_{V_p=a} &= I(a), \nonumber \\
I|_{V_p=b} &= I(b). \label{3}
\end{eqnarray}
A double integration in $V_p$ of equation (\ref{2}) results in
\cite{10}
\begin{equation}
H(V_p) = \int_a^b K\left(V_p,V_p'\right) F\left(V_p'\right) dV_p',
\label{4}
\end{equation}
where
\begin{equation}
\eqalign{H\left(V_p \right) &= I\left(V_p \right) - h\left(V_p
\right),\\
K\left(V_p,V_p'\right) &= \left\{\begin{array}{ccc}
\frac{\left(a-V_p'\right)\left(b-V_p \right)}{b-a},&\; {\rm if}\;
V_p'\leq V_p \\
\frac{\left(a-V_p\right)\left(b-V_p'\right)}{b-a},&\; {\rm if}\;
V_p'> V_p
\end{array}
\right.} \label{5}
\end{equation}
and,
\begin{equation}
h(V_p) = \frac{V_p}{b-a}\left[I(b) - I(a)\right]+
\frac{1}{b-a}\left[bI(a) - aI(b)\right]. \label{6}
\end{equation}
This integral equation is known as the Fredholm equation of first
class and it can be solved numerically by regularization methods.
The Galerkin discretization becomes the more efficient scheme for
this class of problems \cite{11, 12} .

\section{EEDF by the Tikhonov regularization \\
method.}

In order to solve the Fredholm's first class equation (\ref{4}),
the kernel $K$ can be expressed in the form of an expansion in
terms of singular functions $u_i(x)$, $v_i(x)$, and singular
values $k_i$:
\begin{equation}
K(x,y)=\sum_{i=1}^{n}k_i u_i(x) v_i(y). \label{7}
\end{equation}
In the occurrence of symmetry, this reduces itself to an
eigenfunction expansion. It is not necessary that the expansion
should converge pointwise; all we require for an $L_2$ kernel is
that
\begin{equation}
\lim_{n\rightarrow \infty}\int \int \left\{K(x,y) -
\sum_{i=1}^{n}k_i u_i(x) v_i(y)\right\}^2 dx dy=0, \label{8}
\end{equation}
for which it is necessary and sufficient that the series $\sum_i
k_i^2$ converge.

\begin{table}
\caption{\label{t1}Values of constants appearing in (\ref{20}) and
(\ref{21}) obtained by a least-squares fitting.}

\begin{indented}
\lineup
\item[]\begin{tabular}{@{}cccccc}\br
  $\gamma$ & $\beta$ & $V_f$ (V) & $C_0$ & $A_1$ & $A_2$ \cr
\mr
  0.393 & $0.12\times 10^{-3}$ & -13.49 & $0.161\times 10^{-4}$ &
  $0.39\times 10^{-6}$ & $0.131\times 10^{-4}$ \cr
  $A_3$ & $V_s^e$ (V)
  & $u_1$ & $u_2$ & $u_3$ &  \cr
  $0.323\times 10^{-7}$
  & 25.3 & 0.096 & 0.188 & 0.0533 &  \cr
\br
\end{tabular}
\end{indented}
\end{table}

When we have an ill-conditioned matrix $K$, the $k_i$ values
usually becomes very small. In this case the solution of the
system (\ref{4}) proposed by Tikhonov consists in replacing the
ill-posed problem with a stable minimization problem involving a
small positive parameter $\alpha$: instead of attempting to solve
the equation (\ref{4}) directly, we seek to minimize the quadratic
functional \cite{4}
\begin{equation}
\left\Vert Kf-\tilde{h}\right\Vert ^{2}+ \alpha \left \Vert L
f\right\Vert^2 , K\in\Re^{m\times n},m>n,\label{9}
\end{equation}
where $L$ is some linear operator and $\tilde{h}$ denotes
$h+\delta h$. If $L$ is suitably chosen, then the second term has
a smoothing or stabilizing effect on the solution. We may, for
example, take $Lf=f, f'$, or $f''$; if the $k-th$ derivative is
selected, the process is termed $k-th$ order regularization. In
the case $Lf=f$, the solution of the minimization problem
(\ref{10}) is then obtained as the solution of the linear equation
\begin{equation}
\left( K^T K + \alpha I\right)f_\alpha = K^T \tilde{h}. \label{10}
\end{equation}
The operator acting on $f_\alpha$ is clearly positive-definite
when $\alpha>0$ and consequently it has a bounded inverse. Solving
in terms of singular functions we obtain
\begin{equation}
f_\alpha (y) = \sum_i \frac{k_i h_i}{k_i ^2 + \alpha} v_i (y).
\label{11}
\end{equation}
Comparing this with the exact expansion we can see that the effect
of regularization has been to insert a filter factor $k_i/(k_i^2 +
\alpha)$. This is close to unity so long as  $k_i$ is large
compared with $\alpha$ but tends to zero as $k_i \rightarrow 0$,
the rate of transition depending on $\alpha$. If we split
$\tilde{h}$ into $h + \delta h$ the expression (\ref{11}) becomes
\begin{equation}
f_\alpha = \sum_i \frac{k_i h_i}{k_i ^2 + \alpha} v_i (y) +
\frac{k_i \delta h_i}{k_i ^2 + \alpha} v_i (y). \label{12}
\end{equation}
As for the first term, it is advantageous to make $\alpha$ small
in order to reduce the error due to regularization; by contrast,
the second term, which only consists of error, is made small by
taking $\alpha$ large. Thus, there is a conflict, and we would
like to achieve the best compromise \cite{11}.

\begin{table}
\begin{flushleft}
\caption{\label{t2}Values of plasma parameters obtained by the
Langmuir theory, the Tikhonov regularization method and by a
least-squares fitting.}

\begin{indented}
\lineup

\item[]\begin{tabular}{@{}lcccccc}
\br
  & $T_e$ & $n_e$ & $F(\varepsilon)|_{max}$ & $V_s$ & $n_e$ &
  $T_{eff}$ \cr
  & (eV) & (m$^{-3}$) & (m$^{-3}$ eV$^{-1}$) & (V) & (m$^{-3}$) &
  (eV) \cr
  & & & & & [eq. (\ref{22})] & [eq. (\ref{23})] \cr  \mr
1st Derivative & 6.88 & $1.39\times 10^{16}$ & -- & 22.08 & -- &
--  \cr Regularization & 5.66 & $1.54\times 10^{16}$ & $7.34\times
10^{14}$ & --& -- & -- \cr LS fitting & 5.331 &
$1.62\times10^{16}$ & $1.131\times 10^{15}$ & 25.3 & $7.8\times
10^{15}$ & 8.4 \cr \br
\end{tabular}
\end{indented}
\end{flushleft}
\end{table}

\section{Least-squares fitting.}

In the OML theory of ion collection, the ion current flowing to a
negatively biased probe is independent of the shape of the plasma
potential $V(r)$ as long as the current is limited only by the
angular momentum of the orbiting ions \cite{6}. This requires the
arbitrary assumption of either a ``sheath edge" $s$, beyond which
the ion energy distribution is Maxwellian, or a $V(r)$ varying so
slowly that no ``absorption radius", inside of which all ions are
drawn in, exists between the probe and infinity. This condition is
never satisfied even at modest densities. For $s \rightarrow
\infty$ and a Maxwellian ion distribution at temperature $T_i$,
the OML current to a cylinder probe is given by
\begin{eqnarray}
I &= S j_r \left[ \frac{2}{\sqrt{\pi}} \chi^{1/2} + e^\chi
\left(1 -\mbox{erf} (\chi^{1/2})\right) \right] \nonumber \\
&\stackrel{\chi \gg 1}{\longrightarrow} S j_r \frac{2}{\sqrt{\pi}}
\sqrt{1+\chi}, \label{13}
\end{eqnarray}
where $\chi \equiv -eV_p/kTi$ and $j_r$ is the random thermal ion
current. As $T_i \rightarrow 0$, the $T_i$ dependencies of $\chi$
and $j_r$ vanish, and a finite limiting value of the OML current
is reached \cite{13}
\begin{equation}
I \stackrel{T_i \rightarrow 0}{\longrightarrow}A_p
ne\frac{\sqrt{2}}{\pi} \left(\frac{|eV_p|}{M} \right)^{1/2}.
\label{14}
\end{equation}

In the OML theory, the ion current of equation (\ref{14}) can be
represented as
\begin{equation}
I(V_p) = -\beta (V_p-V_f)^\gamma + C_0, \label{15}
\end{equation}
where $\beta$, $\gamma$, $V_f$ and $C_0$ are constants to be
fitted. In our case $V_f$ acquires the meaning of the float
potential.

The three trial functions in the least-squares fitting of the I-V
characteristic for the electronic component have the form
\begin{equation}
F_1(V_p) = A_1 (V_p - V_s) e^{-u_1(V_p - V_s)}, \label{16}
\end{equation}
and
\begin{equation}
F_i(V_p) = A_i (V_p -V_s)^2 e^{-[u_i(V_p -V_s)]^2}, \;\;
(i=2,3)\label{17}
\end{equation}
where $A_i$, $u_i, \; (i=2,3)$ and $V_s$ are constants that must
be valued. Here $V_s$ adopts the character of the plasma
potential.

Substituting equation (\ref{16}) into an expression for the
current to the probe, written in the form \cite{13}
\begin{equation}
I\left(V\right)=B_1 \int_{x}^{\infty} (V_p-x)F(V_p)dV_p,
\label{18}
\end{equation}
where $B_1 = (4/eS)\sqrt{1 /2m_e}$, $x=eV_p$ and in place of
$F(V_p)$ we take any of functions (\ref{16})-(\ref{17}), we obtain
for the electronic component
\begin{equation}
\frac{I_1(V_p)}{B_1} = \frac{A_1}{u_1} \left[ 2 + u_1\left( V_p -
V_s\right) \right] \exp\left[-u_1 \left( V_p - V_s\right) \right],
\label{19}
\end{equation}
where the subindex $1$ in $I$ refers to trial function (\ref{16}).
Analogously, from relations (\ref{17}) we get
\begin{eqnarray}
\fl \frac{I_i(V_p)}{B_1} = \frac{A_i}{4u_i^2} \left\{ e
\exp\left[-u_i^2\left( V_p - V_s\right)^2\right] \right.\nonumber \\
\lo+ \left. u_i \sqrt{\pi} \left(V_p - V_s\right) \left(
\mbox{Erf} \left[ u_i \left( V_p - V_s \right) \right] - 1 \right)
\right\} \; \; (i=2,3) \label{20}
\end{eqnarray}

\section{Analysis and discussion.}

With a view to obtaining accurate values of $n$, $T_e$ and $V_s$
(but not $T_i$) from the EEDF, we shall illustrate a procedure
using data obtained in an ECR discharge with $P= 10$ mTorr in
argon gas, taken with an rf-compensated cylindrical probe with a
radius 0.4 mm and  4 mm in length.

It is considered that the current is collected within the area at
the tip of the probe expressed by $S=2\pi R_p L + \pi R_p^2$,
where $L$ is the probe length. The entire I-V curve in this
example is shown in \fref{F1}.

From the Langmuir theory, assuming a Maxwellian distribution for
electrons \cite{13}, we obtain in the case of the $I-V$ curve of
\fref{F1}, that $T_e = 6.88$ eV, $V_s = 22.078$ V, and $n_e = 1.39
\times 10^{16}$ m$^{-3}$. Thus, the first derivative shown in
\fref{F1} as well as experimental data are used. Here, it is
important to notice that the extremum of the first derivative is
not completely well defined, as observed in \fref{F1}.

The EEDF is calculated following the procedure described in \S 3,
by the integration of equation (\ref{4}) with kernel (\ref{5}),
and using the Galerkin discretization method \cite{14}.

In \fref{F3} the EEDF obtained by the Tikhonov technique of zeroth
order is shown, along with the corresponding values of the
regularization parameters $\alpha$ = (0.09, 0.08, 0.07, 0.06,
0.05, 0.04, 0.03, 0.02).

From these charts we gather that the most probable energy is $T_e=
5.66$ eV. With this electron temperature, it is easy to calculate
the plasma electron density by the relation
\begin{equation}
I(V_p)|_{V_p=V_s} = eS n_e \left( \frac{T_e}{2\pi
m_e}\right)^{1/2}, \label{21}
\end{equation}
obtaining that $n_e = 1.54\times 10^{16}$ m$^{-3}$. It is also
important to notice that the maximum value of the EEDF by the
Tikhonov regularization method takes the value of
$F(\varepsilon)|_{max}=7.34\times 10^{14}$ m$^{-3}$ eV$^{-1}$.
After an analysis of several calculations, we can say that the
bumps observed in the EEDF are related with numerical effects and
are originated by the irregularities of the I-V curve.

These results are validated in the following by a least squares
fitting. As it has been described in \S IV about the ionic part of
the I-V characteristic, we fit the data to equation (\ref{16})
determining the values for the constants $\gamma$, $\beta$ and
$C_0$. These values are summarized in Table~\ref{t1}. This fitting
is shown in \fref{F1}. In \fref{F2}, the measured values $I^2$
versus $V$ curves are plotted \cite{13}. From these charts, we can
observe that the obtained values for the constants show a good
agreement with the data values.

\begin{table}
\caption{\label{t3}Values of plasma parameters obtained by the
Langmuir theory, the Tikhonov regularization method and by a
least-squares fitting for the energetic group of electrons.}

\begin{indented}
\lineup

\item[]\begin{tabular}{@{}lccc}
\br
  & $T_e$ & $n_e$ (V) & $F(\varepsilon)|_{max}$  \cr
  & (eV) & (m$^{-3}$) & (m$^{-3}$ eV$^{-1}$)   \cr \mr
  Regularization & 21.93 & $8.02 \times 10^{15}$ & $3.31 \times
  10^{14}$ \cr
  LS fitting & 18.831 & $8.66 \times 10^{15}$ & $3.42 \times
  10^{14}$  \cr \br
\end{tabular}
\end{indented}
\end{table}

Analogously, for the electronic component in the I-V curve,  we
now employ expressions (\ref{18}) and (\ref{19}) for a least
squares fit to the data. The corresponding values of constants
appearing in (\ref{18}) and (\ref{19}) are given in
Table~\ref{t1}. Then, we can express the EEDF as a sum of
functions (\ref{16})-(\ref{17}). This distribution is plotted in
\fref{F1}. Introducing the corresponding dimensions, we obtain a
comparison with the EEDF obtained by the Tikhonov regularization
method shown in \fref{F3}. For the energetic group of electrons we
obtained that $T_e = 18.831$ eV, $n_e = 8.66E15$ m$^{-3}$, and
$F(\varepsilon)|_{max} = 3.42E13$ m$^{-3}$ eV$^{-1}$. From the
least-squares distribution we obtain for the main group of
electrons that the most probable energy is $T_e = 5.331$ eV. The
maximum value found for the EEDF is $F(\varepsilon)|_{max} =
1.131\times 10^{15}$ m$^{-3}$ eV$^{-1}$. Considering this value
for the temperature $T_e$, we find by equation (\ref{21}) the
density $n_e = 1.62\times 10^{16}$ m$^{-3}$. For the values
computed after fitting, it is possible to accomplish the
integration of the adjusted characteristic, to obtain \cite{15,
16}
\begin{equation}
n_e = \int_{0}^{\infty} F(\varepsilon) d\varepsilon = 7.8\times
10^{15} \; \mbox{m}^{-3}, \label{22}
\end{equation}
and
\begin{equation}
T_{eff} = \frac{2}{3} \langle \varepsilon \rangle = 2(3n_e)^{-1}
\int_0^\infty \varepsilon F(\varepsilon) d\varepsilon = 8.40 \;
\mbox{eV}. \label{23}
\end{equation}

Here, the  related problem of the indetermination of the zero
position in the EEDF, as it is observed from \fref{F3}, deserves
some attention. This problem will always  be present when no
electronic saturation is reached. Several results obtained by
applying the two methods are summarized in Table~\ref{t2} and
Table~\ref{t3}.

\ack

This work was partially supported by CONACyT, Mexico, under
contract 33873-E.

\Figures
\begin{figure}
\caption{\label{F1}Sample I-V curve to be analyzed obtained from
an 0.4 mm diam, 4 mm long probe in an ECR discharge in 10 mTorr
argon gas (solid). First derivative of the I-V curve (dotted), and
saturation ion and electronic currents computed after fitting
(dashed).}
\end{figure}

\begin{figure}
\caption{\label{F2}Square of saturation ion current vs probe
voltage as measured (solid) and as computed after fitting
(dashed).}
\end{figure}

\begin{figure}
\caption{\label{F3}EEDF resulting from the least squares fitting
(solid) and by the Tikhonov regularization method (dotted). Here
are shone the locations of the main and energetic groups of
electrons respecting the energy.}
\end{figure}

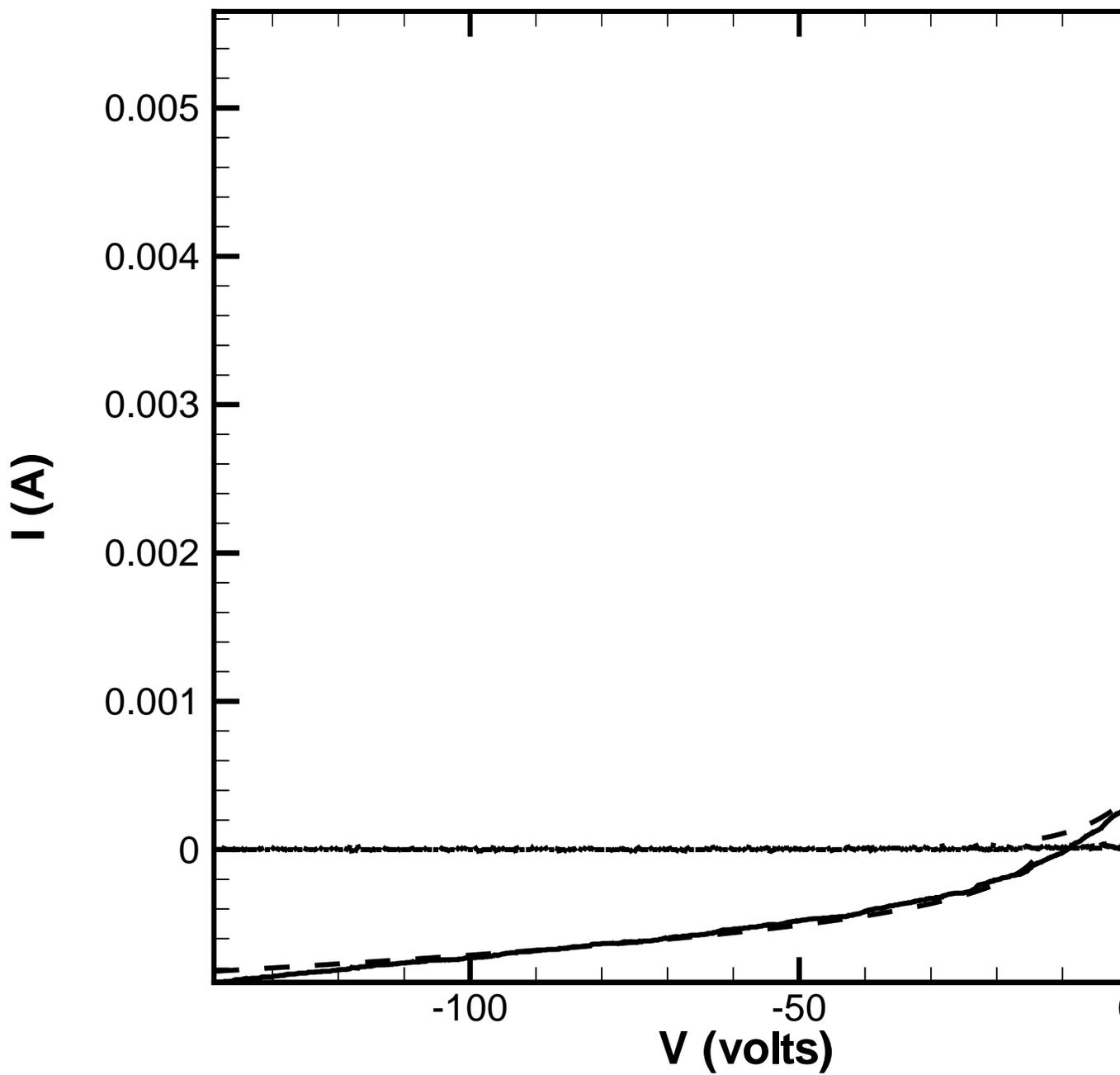
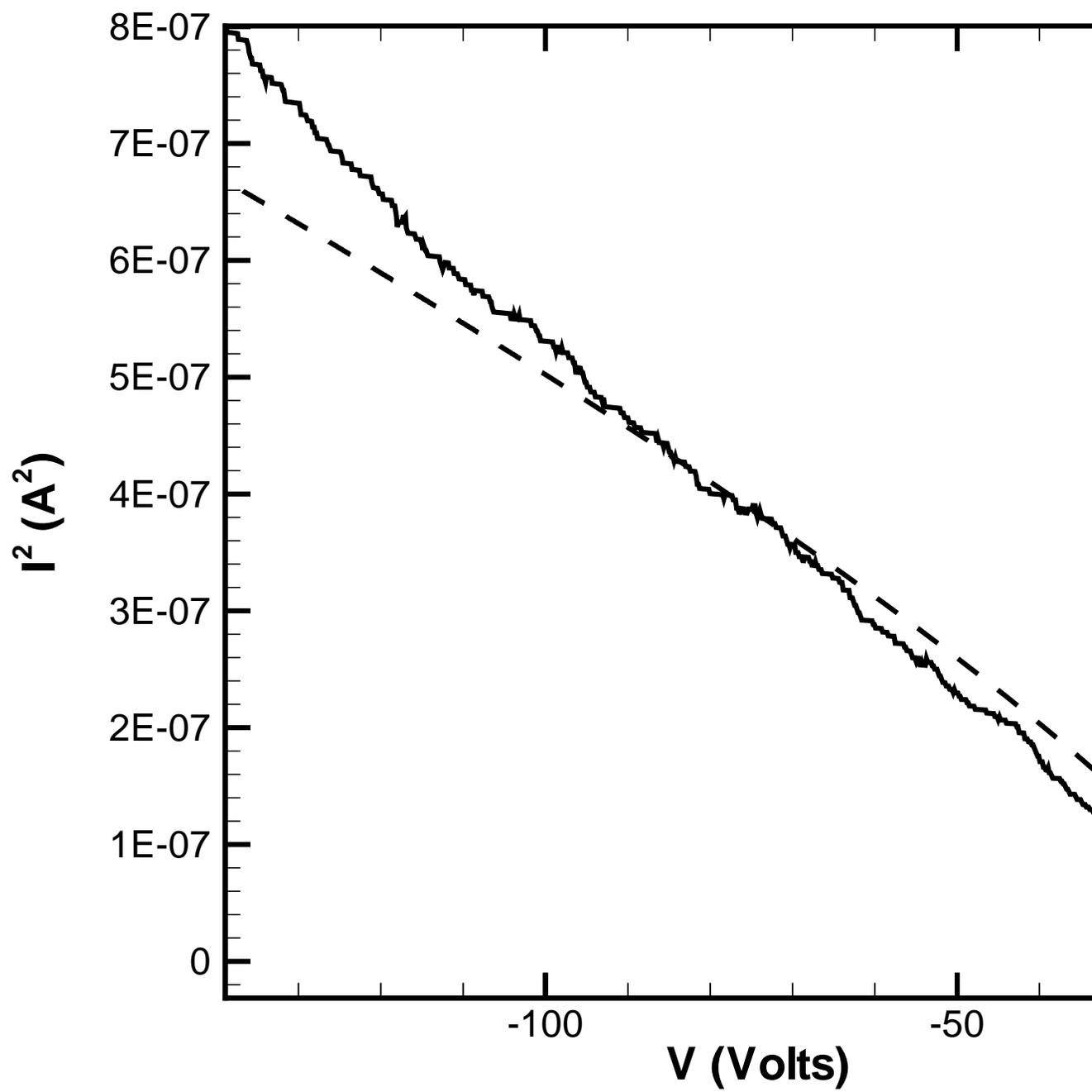
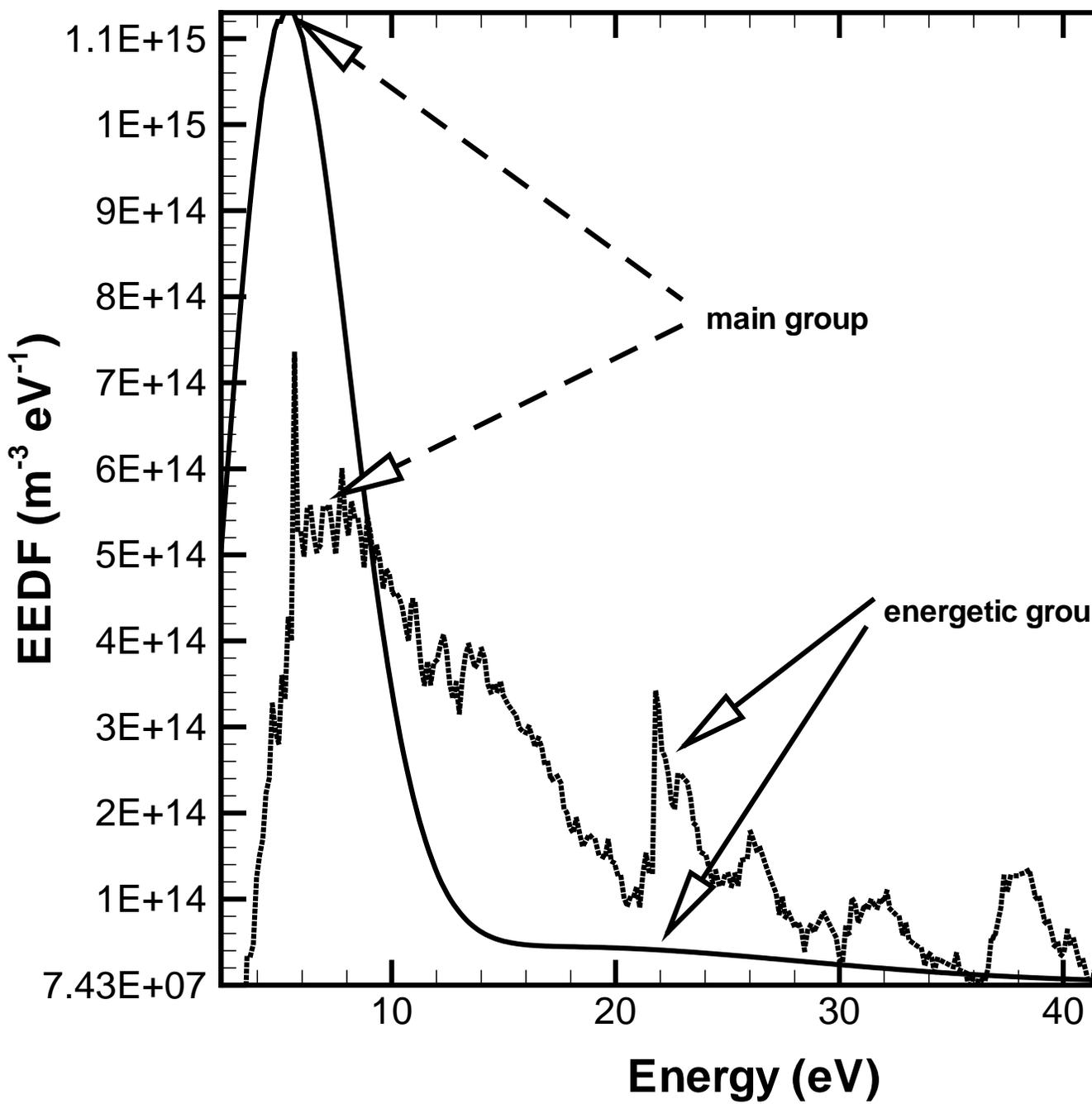
\end{document}